\documentclass{ws-procs9x6}

\begin{document}

\title{The Status of Core-collapse Supernova Simulations}

\author{CHRISTIAN~Y. CARDALL}

\address{Physics Division, \\
        Oak Ridge National Laboratory, \\
        Oak Ridge, TN 37831-6354, USA, \\
        E-mail: ccardall@mail.phy.ornl.gov}
\address{Department of Physics and Astronomy, \\
         University of Tennessee, \\
	Knoxville, TN 37996-1200, USA} 
\address{Joint Institute for Heavy Ion Research, \\
        Oak Ridge National Laboratory, \\
        Oak Ridge, TN 37831-6374, USA}


\maketitle

\abstracts{
Core-collapse supernovae can be used to place limits on
dark matter candidate particles, but the strength of these
limits depends on the depth of our theoretical understanding 
of these astrophysical events. To date, limitations on computing
power have prevented inclusion of all the physics
that would constitute a realistic simulation. The TeraScale
Supernova Initiative (TSI) will overcome these obstacles in the
next few years, elucidating the explosion mechanism and
other phenomena closely associated with the core collapse
of massive stars.   
}

\section{What are supernovae?}

The term ``supernova'' dates from the early 1930s, but the concept 
was around in the 1920s. With the realization that the spiral nebulae 
are separate galaxies comparable to our Milky Way, it was recognized 
that the ``novae'' or ``new stars'' seen in these nebulae would have to be much more luminous than typical novae occuring in our galaxy. 
The phrases 
``giant novae'', novae of ``impossibly great absolute magnitudes'', 
``exceptional novae'', and the German term 
``Hauptnovae'' or ``chief novae'' were used during the 
1920s.\cite{osterbrock01}
In a review article Zwicky explained that it was deduced that ``supernovae'' 
were about a thousand times as luminous as ``common novae'', and 
claimed that ``Baade and I first introduced the term 'supernovae' in seminars and in a lecture course on astrophysics at the California Institute of Technology in 1931.''\cite{zwicky40}

Supernovae are classified by astronomers into two broad classes based on their spectra.\cite{filippenko97}
These classes are ``Type I'', which have no hydrogen features, and 
``Type II'', which have obvious hydrogen features. These types have further 
subcategories, depending on the presence or absence of silicon and 
helium features in Type I, and the presence or absence of narrow 
hydrogen features in the case of Type II. In particular, 
supernovae of Type Ia exhibit strong silicon lines, 
those of Type Ib have helium lines, and those of Type Ic do not have 
either of these. Astronomers have also identified a number of distinct
characteristics in supernova light curves
(total luminosity as a function of time). 

There are two basic physical mechanisms for supernovae, but these do not line up cleanly with the observational categories of Type I and Type II. Type Ia supernovae are caused by accretion of matter from a companion star onto a white dwarf, which induces a thermonuclear runaway that consumes the entire white dwarf.
Supernovae of Type Ib, Ic, and II are produced by a totally different mechanism: the catastropic collapse of the core of a massive star. The observational distinctions of presence or absence of hydrogen or helium turn out to be unrelated to the mechanism; they depend on whether the outer hydrogen and helium layers of the star, which have nothing to do with the collapsing core, have been lost to winds or accretion onto a companion during stellar evolution. Of the two
physical mechanisms, core-collapse supernovae are the focus of the
present discussion.

For most of their existence stars burn hydrogen into helium. 
In massive stars, temperatures and densities become sufficiently high to burn 
to carbon, oxygen, neon, magnesium, and silicon and iron group elements. 
The iron group nuclei are the most tightly bound, and here burning in 
the core ceases. The iron core---supported by electron 
degeneracy pressure---eventually 
becomes unstable. Its inner portion undergoes homologous collapse
(velocity proportional to radius), and the outer portion collapses 
supersonically. 
Electron capture on nuclei is one instability leading to collapse, and 
this process continues throughout collapse, releasing free-streaming
neutrinos 
until densities and temperatures become so high that even neutrinos are 
trapped. Collapse is halted when the matter reaches nuclear density; 
at this point a shock wave forms at the boundary between the homologous
and supersonically collapsing regions. The shock begins to move out,
and after the shock passes
some distance beyond the surface of the newly born neutron star, 
it stalls as energy 
is lost to neutrino emission and dissociation of infalling heavy 
nuclei falling through the shock.

The nascent neutron star is a hot thermal bath of dense nuclear matter, 
electron/positron pairs, photons, and neutrinos, containing most of 
the gravitational potential energy released during core collapse. 
Neutrinos, having the weakest interactions, are the most efficient 
means of cooling; they diffuse outward on a time scale of seconds 
towards a semi-transparent region near the surface of the neutron star, 
and eventually escape with about 99\% of the released gravitational energy.

In the reigning paradigm---the neutrino-driven explosion 
mechanism\cite{bethe85}---the supernova explosion is launched as a result of 
neutrino heating of the 
material behind the 
stalled shock, resulting in the revival of the shock and 
its propulsion through the outer layers of the star. 
This process may be aided by convection in two regions. 
First, loss of electron neutrinos from the outer layers of the neutron star 
causes composition gradients that drive convection, 
which boosts neutrino luminosities by bringing hotter material to the surface. 
Second, heating decreases away from the neutron star surface, 
giving rise to a negative entropy gradient. 
The resulting convection increases the efficiency of neutrino heating 
by delivering heated material to the region just behind the shock.

As the neutrinos are transported from inside the neutron star, they go from a nearly isotropic diffusive regime to a strongly forward-peaked free-streaming region. Modeling this process accurately requires tracking both the energy and angle dependence of the neutrino distribution functions.

\section{Survey of core-collapse simulations}

Supernovae have a rich phenomenology---observations of many types that 
modelers would like to reproduce and explain. 
Chief among these is the explosion itself, which is not produced 
robustly and convincingly in simulations. 
As mentioned previously, 99\% of the gravitational 
potential energy released during collapse escapes as neutrinos; in 
comparison, 
the kinetic energy of expelled matter accounts for about 1\%, 
and the optical display is just a fraction of this. 
Energetically, supernovae are essentially neutrino events; 
the explosion is just a minor sideshow, the optical display a trivial detail. 
That the explosion is such a minor part of the system is what makes it 
so challenging to model convincingly. But the optical data are what 
we perceive with our unaided inborn detectors---our eyes---and in our 
anthropic chauvinism, explaining the explosion seems most interesting.

While the explosion is of obvious interest, neutrino signatures are also 
of great importance. The handful of neutrinos detected from supernova 
SN1987A confirmed theoretical predictions of neutrinos releasing 
the gravitational energy on a time scale of seconds. This was a remarkable 
success of supernova theory and modeling. The neutrinos are also 
important because their detection allows limits to be placed on dark matter 
candidates, such as axions and sterile neutrinos.

There are many other interesting observables, including pulsar spins,
kick velocities, and magnetic fields; gravitational waves; element
abundances; and all kinds of measurements across the electromagnetic
spectrum. Core-collapse simulations---the subject of the present
discussion---typically 
address the explosion mechanism, neutrino signatures, remnant pulsar
properties,
and gravitational waves. Another class of simulation---not discussed 
here---{\em assumes} 
a successful explosion and studies the interaction of the shock
with the surrounding layers of the star (and beyond) in order to 
study things like nucleosynthesis 
and measurements across the electromagnetic spectrum.

From the description of the core-collapse supernova process in the
previous section, several key aspects of physics that a simulation must 
address can be identified:

{\em Neutrino transport/interactions:} Because neutrinos dominate the system, their treatment is very important, including the number of spatial dimensions treated; dependence on both energy and angle in order to properly model the transition from isotropic diffusion to forward-peaked free streaming; relativistic effects; and comprehensiveness of interactions.

{\em Hydrodynamics/gravitation:} Convection---both inside and outside the
nascent neutron star---can play an important role, so allowance for flows in multiple spatial dimensions is important in the hydrodynamics. The newly born neutron is sufficiently compact that a general relativistic description should 
 ultimately be included.

{\em Equation of state/composition:} Determination of realistic equations
of state of dense nuclear matter at finite temperature involves cutting-edge
nuclear physics, as does the determination of neutrino interaction rates
with the variety of nuclear species encountered in the supernova environment.

{\em Diagnostics:} Very important to making convincing explosions 
is fastidious accounting of total lepton number and energy. Because 
the explosion energy is only 1\% of the basic energy scale in the problem, 
a determination of the explosion energy accurate to 10\% requires that
total energy be conserved at a level of about one part in $10^8$ per time step
(allowing for systematic error accrual over $\sim 10^5$ time steps).

Simulations of collapse and bounce have been performed by many groups. In 
briefly describing this work, I list (in alphabetical order) the institutions 
that represent the ``centers of gravity'' of many of these groups:

{\em Livermore National Laboratory:} The neutrino transport in these
simulations\cite{bowers82,bowers82b,mayle88,wilson93}
was energy-dependent, included some relativistic effects, and
had a decent set of neutrino interactions; however, the transport was
spatially one-dimensional (spherically symmetric). The hydrodynamics and
gravitation, while relativistic, were also spherically symmetric. 
Multidimensional effects were mocked up with a mixing-length approach, and
without this explosions were not seen in these models. Inclusion of several
nuclear species in a burning network was a high point of these simulations.
Explosions were seen in these models, but there was no published 
accounting of lepton number and energy conservation.

{\em Los Alamos National Laboratory:} Many groups published spatially
two-dimensional simulations in the early 1990s, and a group centered at 
Los Alamos was one of the first.\cite{herant94} 
Descended from those efforts was a recent simulation in three spatial
dimensions by Fryer and Warren.\cite{fryer02} 
The high point of these simulations was their three-dimensional hydrodynamics, 
and some relativistic effects were included
in both the neutrino transport and gravitation.
An important liability, however, was the crude treatment 
of neutrino transport, in which dependence on both energy 
and angle were integrated out, and some important interactions 
were left out. Explosions were seen in these models, but
there was no published accounting of lepton 
number and energy conservation. 

{\em Max Planck Institute for Astrophysics:} This group also performed 
simulations with two-dimensional hydrodynamics in the mid 1990s. Separate simulations were performed for the nascent neutron star\cite{keil96} 
and the region above the neutron star.\cite{janka96} 
In the latter simulations the neutrino luminosities were 
parametrized, and explosions were seen if these luminosities were set high enough. While some relativistic effects were included in both the hydrodynamics 
and neutrino fields, these were parametrized models with no serious neutrino
transport. There was no detailed accounting of lepton number and energy
conservation. 

In more recent work the Max Planck group has published studies in 
spherical symmetry, but with sophisticated neutrino 
transport.\cite{rampp00,rampp02} 
The neutrino transport is dependent on both energy and angle, and includes
some relativistic effects. This group has also considered a full range of
neutrino interactions.\cite{buras02} 
Important limitations of results published to date were the restriction to
spherical symmetry and Newtonian hydrodynamics, but initial results on
multidimensional models with sophisticated neutrino transport 
and approximate relativity have recently been
reported.\cite{janka02,janka02b}
While some attention to the accounting of lepton number and energy was
reported in connection with test problems, 
it was not reported in detail in connection with
full simulations. No explosions were seen in simulations with the most
comprehensive treatments of neutrino transport.

{\em Oak Ridge National Laboratory:} Like the group at
Max Planck Institute for Astrophysics, this group published 
separate studies of the nascent neutron star\cite{mezzacappa98} 
and the neutrino-heated
region\cite{mezzacappa98b} with two-dimensional hydrodynamics. 
The neutrino transport included
some relativistic effects, retained the energy dependence of the
neutrino distributions, and took some care regarding the conservation
of energy and lepton number. However, these neutrino distributions were taken
from a spherically symmetric simulation and imposed onto the 
two-dimensional hydrodynamics. In addition, the hydrodynamics was
nonrelativistic. Unlike other multidimensional models published in the 1990s,
no explosions were seen in these simulations.

\begin{figure}[ht]
\centerline{\epsfxsize=3.6in\epsfbox{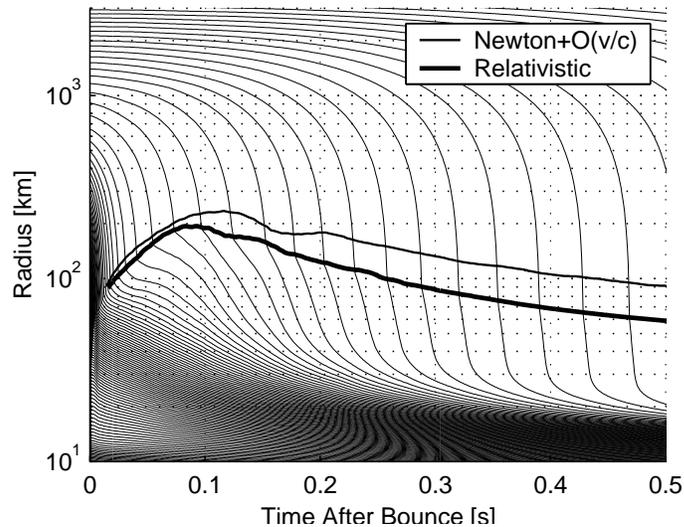}}   
\caption{Failure of spherical model to explode. Thin
lines:  mass shell trajectories.  Thicker lines: shock trajectories
in a Newtonian hydrodynamics, $O(v/c)$ neutrino transport model
and a fully relativistic model.}
\label{traces}
\end{figure}

More recently the Oak Ridge group has produced simulations in spherical 
symmetry but with sophisticated neutrino 
transport.\cite{mezzacappa01,liebendoerfer01b,liebendoerfer02} These 
recent simulations had some notable high points. 
They included realistic neutrino transport, tracking both the energy and 
angle dependence of the neutrinos. 
Unlike other core-collapse simulations, 
they were fully relativistic in both the transport---including redshifts 
and trajectory bending---and in the hydrodynamics/gravitation.
The price that has been paid for these advances is that the models were 
spherically symmetric, and therefore unrealistic in that respect.
Another high point of these simulations was the careful attention paid 
to energy and lepton number conservation.\cite{liebendoerfer02} 
Great effort went into 
assuring that the finite difference representations of the 
partial differential equations were consistent with number and 
energy conservation. This is a standard all simulations eventually 
must meet in order to be truly credible with respect to conclusions about the 
explosion mechanism. In this case the simulations provided convincing 
evidence that explosions simply do not occur in spherical 
symmetry (at least with standard neutrino physics). This is indicated
in Figure \ref{traces}, which shows the initial outward motion of the
shock and its subsequent stagnation and infall.\cite{liebendoerfer01b}

{\em University of Arizona:} This group also performed simulations
with multidimensional hydrodynamics in the mid 1990s.\cite{burrows95} 
The neutrino 
transport was integrated over both energy and angles, impairing the 
realism of the models. They were also nonrelativistic. 
While these
were exploding models, there was no detailed accounting of energy and
lepton number conservation.\footnote{Very recently this group reported simulations with Newtonian
hydrodynamics and sophisticated neutrino transport in spherical
symmetry; no explosions were seen.\cite{thompson02}}
 

\section{The TeraScale Supernova Initiative}

The overview of core-collapse supernova simulations presented in the
last section demonstrates a fundamental trade-off required by the computing
power available during the past decade: A sadistic choice between 
multidimensional hydrodynamics and spatially multidimensional, 
energy- and angle-dependent neutrino
transport, two non-negotiable aspects of realism.
As the new millenium begins, computing power has advanced to the point
that neither of these crucial pieces of physics need be sacrificed, as
initial results\cite{janka02,janka02b} from 
the group at the Max Planck Institute for Astrophysics indicates.
Another program designed to take core-collapse supernova modeling to the
next level is the TeraScale Supernova Initiative. 

The TeraScale Supernova Initiative\footnote{\sf http://www.phy.ornl.gov/tsi}
(TSI) is a large collaboration funded by the U.S. 
Department of Energy for several years with the mission to explain 
the supernova phenomena most closely associated with core collapse: 
the successful launch of the shock (i.e. understand the explosion 
mechanism); neutrino signatures; pulsar spins, kick velocities, and 
magnetic fields; gravitational waves; and heavy element 
($r-$process) abundances.

This initiative---which grows out of past efforts of the supernova modeling
group at Oak Ridge National Laboratory---has a diverse and 
experienced investigator team, including some 40 investigators
from 12 institutions.\footnote{Inclusion of collaborators who
are not members of TSI brings the total number of people involved
to 99, from 28 institutions.} These investigators include people whose 
life-long work has been in the areas needed in supernova science, 
including radiation transport, magneto-hydrodynamics, nuclear and 
weak interaction physics, and needed aspects of computer science. 

TSI also has the support of the U.S. Department of Energy's computational 
infrastructure. This includes high priority on the DOE's terascale 
machines---which feature several $10^{12}$ bytes of memory and 
floating point operations per second---and access to the expertise of 
teams specializing in various aspects of 
high-performance computing, including advanced solver algorithms,
computational meshes, performance on parallel architectures, data
management and visualization, and software reusability and interoperability.

Some recent science from TSI involves pure hydrodynamics simulations 
(no neutrino transport).\cite{blondin02} 
A standing accretion shock is an analytic 
solution in spherical symmetry. Its parameters are matched to the 
density profile in a realistic simulation when the shock stalls, 
and this is used as an initial condition for two- and three-dimensional 
simulations. It turns out that the standing accretion 
shock is unstable in two and three dimensions to the lowest order 
modes; the average shock radius and turbulent energy increase steadily 
with time at the expense of thermal and gravitational energy of the 
gas.\footnote{Movies
available at {\sf http://www.phy.ornl.gov/tsi/pages/simulations.html}.}
The mechanism of instability is nonlinear feedback between aspherical
pressure waves rising from small radii and regions
characterized by transverse flow velocities---generated by 
asphericities in the shock---that advect inwards.
This must be studied in more detail in simulations involving
neutrino transport and realistic equations of state, but 
this hydrodynamic instability may play an important role in the 
supernova mechanism and provide an explanation for aspect ratios $\sim 2$
inferred from spectropolarimetry data.\cite{hoeflich01}

\section{Summary and Implications for Dark Matter}

While many simulations have been performed over the years, it
cannot yet be claimed that the supernova explosion mechanism
is understood. Models with energy- and angle-dependent neutrino
transport have been studied in spherical symmetry, and explosions
have not been seen. Many (though not all) models with 
multidimensional hydrodynamics do exhibit explosions, but these
have employed neutrino transport that is too crude to make 
firm conclusions about the explosion mechanism. Computing power
has advanced to the point that models with {\em both} sophisticated
neutrino transport and multidimensional hydrodynamics are within
reach; the TeraScale Supernova Initiative (TSI) is one effort
underway to perform such simulations.

Finally, in conclusion, a word on the subject of the conference:
dark matter. The nascent neutron star is a
hot (temperature of order 50 MeV) and dense (baryon mass density of 
order $10^{14}$ g cm$^{-3}$) environment. Should hypothetical particles like
the axion (a cold dark matter candidate) or a keV-mass sterile neutrino
(a warm dark matter candidate) exist, they could be produced in
the extreme conditions present in the newly born neutron star.
However, copious production and emission of such weakly coupled 
particles would cause the neutron star to cool more quickly than 
it would if neutrinos were fastest means of cooling. 
The handful of neutrinos detected from supernova SN1987A
confirms the basic theoretical understanding of stellar core collapse,
with associated trapping of neutrinos and their subsequent emission
on a time scale of several seconds; this allows limits to be placed
on the coupling strength of hypothetical dark matter particles
like the axion\cite{raffelt99} and sterile neutrino.\cite{abazajian01}
Presently, these limits are of necessity rather conservative, due to a lack
of detailed understanding of the explosion mechanism and subsequent 
uncertainties about precise conditions in the nascent neutron star.
A new generation of simulations---such as those being 
pursued by TSI---promises 
to reveal the explosion mechanism and paint a detailed 
and realistic picture 
of the physical conditions in the hot and dense neutron star,
providing a basis for strengthened limits on the properties of dark
matter candidate particles.

\section*{Acknowledgments}
\uppercase{T}his work was supported 
by the \uppercase{U.S. D}epartment of \uppercase{E}nergy (\uppercase{D}o\uppercase{E}) \uppercase{HENP S}cientific \uppercase{D}iscovery \uppercase{T}hrough
\uppercase{A}dvanced \uppercase{C}omputing \uppercase{P}rogram; \uppercase{O}ak \uppercase{R}idge \uppercase{N}ational \uppercase{L}aboratory, managed by \uppercase{UT-B}attelle, \uppercase{LLC}, for the \uppercase{D}o\uppercase{E} under contract \uppercase{DE-AC05-00OR22725}; the \uppercase{J}oint \uppercase{I}nstitute
for \uppercase{H}eavy \uppercase{I}on \uppercase{R}esearch; and a \uppercase{D}o\uppercase{E PECASE} grant.


\begin{thebibliography}{0}

\bibitem{osterbrock01} D. E. Osterbrock, {\it American Astronomical 
  Society Meeting} {\bf 199}, {\#}15.01 (2001).

\bibitem{zwicky40} F. Zwicky, {\it Rev. Mod. Phys.} {\bf 12}, 66 (1940).

\bibitem{filippenko97} A. V. Filippenko, 
  {\it Ann. Rev. Astron. Astrophys.} {\bf 35}, 309 (1997).

\bibitem{bethe85} H. A. Bethe and J. R.  Wilson,
  {\it Astrophys. J.} {\bf 295}, 14 (1985).

\bibitem{bowers82}  R.~L. Bowers and J.~R. Wilson,
    {\it Astrophys. J. Suppl.} {\bf 50}, 115 (1982).

\bibitem{bowers82b} R. Bowers and J.~R. Wilson,
  {\it Astrophys. J.} {\bf 263}, 366 (1982).

\bibitem{mayle88} R. Mayle and J.~R. Wilson,
    {\it Astrophys. J.} {\bf 334}, 909 (1988).

\bibitem{wilson93} J.~R. Wilson and R.~W. Mayle,
 {\it Phys. Rep.} {\bf 227}, 97 (1993).

\bibitem{herant94} M. Herant, W. Benz, W.~R. Hix, C.~L. Fryer, and
  S.~A. Colgate, {\it Astrophys. J.} {\bf 435}, 339 (1994).

\bibitem{fryer02} C.~L. Fryer and M.~S. Warren,
    {\it Astrophys. J. Lett.} {\bf 574}, 65 (2002).

\bibitem{keil96}  W. Keil, H.-T. Janka,  and E. Mueller,
    {\it Astrophys. J. Lett.} {\bf  473}, 111 (1996).

\bibitem{janka96} H.-T. Janka and E. Mueller,
    {\it Astron. Astrophys.} {\bf 306}, 167 (1996).

\bibitem{rampp00} M. Rampp and H.-T. Janka, 
  {\it Astrophys. J. Lett.} {\bf 539}, 33 (2000).

\bibitem{rampp02} M. Rampp and  H.-T. Janka,
  astro-ph/0203101.

\bibitem{buras02} R. Buras, H.-T. Janka, M. T. Keil, G. G. Raffelt, 
     and M. Rampp, astro-ph/0205006.

\bibitem{janka02} H.-T. Janka, R. Buras, and M. Rampp,
  in {\it Proceedingsof the 7th International Symposium on 
  Nuclei in the Cosmos}, astro-ph/0212317. 

\bibitem{janka02b} H.-T. Janka, R. Buras, K. Kifonidis, T. Plewa, and
  M. Rampp, in {\it Proceedings of the ESO/MPA/MPE Workshop ``From 
  Twilight to Highlight''}, astro-ph/0212316.

\bibitem{mezzacappa98} A. Mezzacappa, A.~C.Calder, S.~W. Bruenn, 
        J.~M. Blondin,  M.~W. Guidry, M.~R. Strayer,  and 
        A.~S. Umar,
  {\it Astrophys. J.} {\bf 493}, 848 (1998).

\bibitem{mezzacappa98b} A. Mezzacappa, A.~C.Calder, S.~W. Bruenn, 
        J.~M. Blondin,  M.~W. Guidry, M.~R. Strayer,  and 
        A.~S. Umar,
  {\it Astrophys. J.} {\bf 495}, 911 (1998).

\bibitem{mezzacappa01} A. Mezzacappa, M. Liebend\"orfer, O.~E.~B. Messer,
  W.~R. Hix, F.-K. Thielemann, and S.~W. Bruenn, 
  {\it Phys. Rev. Lett.} {\bf 86}, 1935 (2001).

\bibitem{liebendoerfer01b} M. Liebend{\" o}rfer, A. Mezzacappa, F. 
  Thielemann, O.~E. Messer,  W.~R. Hix, and S.~W. Bruenn,
  {\it Phys. Rev.} {\bf D63}, 103004 (2001).

\bibitem{liebendoerfer02} M. Liebend\"orfer, O.~E.~B. Messer, 
                  A. Mezzacappa, S.~W. Bruenn, C.~Y. Cardall, 
                  and F.-K. Thielemann,
 astro-ph/0207036.

\bibitem{burrows95} A. Burrows, J. Hayes, and B.~A. Fryxell, 
  {\it Astrophys. J.} {\bf 450}, 830 (1995).

\bibitem{thompson02} T.~A. Thompson, A. Burrows, and P.~A. Pinto,
   astro-ph/0211194.

\bibitem{blondin02} J. M. Blondin, A. Mezzacappa, and C. DeMarino,
{\it Astrophys. J.} in press, astro-ph/0210634.

\bibitem{hoeflich01} P. H{\" o}flich, A. Khovkhlov, and L. Wang,
  in {\it 20th Texas Symposium on Relativistic Astrophysics}, 459 (2002).

\bibitem{raffelt99} G. G. Raffelt,
  {\it Ann. Rev.Nucl. Part. Sci.} {\bf 49}, 163 (1999).

\bibitem{abazajian01} K. Abazajian, G. M. Fuller, and M. Patel, 
  {\it Phys. Rev.} {\bf D64}, 023501 (2001).

\end{thebibliography}
\end{document}